\documentclass[12pt]{iopart}
\usepackage{graphicx}% Include figure files
\usepackage{epsf}
\usepackage{amssymb}
\usepackage{array}
\begin{document}

\title[]{Dynamics of ghost domains in spin-glasses\footnote{\small To appear in special issue of Journal of Physics A entitled "Statistical Physics of Disordered Systems: from Real Materials to Optimization and Codes" }.}

\author{Hajime Yoshino
\footnote[3]{E-mail: yoshino@ess.sci.osaka-u.ac.jp}
}

\address{Department of Earth and Space Science, Faculty of Science,
Osaka University, Toyonaka, 560-0043 Osaka, Japan}

\begin{abstract}
We revisit the problem of how spin-glasses ``heal'' after being 
exposed to tortuous perturbations by the temperature/bond 
chaos effects in temperature/bond cycling protocols.
Revised scaling arguments suggest the amplitude
of the order parameter within ghost domains recovers very slowly
as compared with the rate it is reduced by the strong perturbations.
The parallel evolution of the order parameter 
and the size of the ghost domains can be examined 
in simulations and experiments by measurements of a 
memory auto-correlation function which exhibits a ``memory peak'' 
at the time scale of the age imprinted in the ghost domains.
These expectations are confirmed by Monte Calro simulations of
an Edwards-Anderson Ising spin-glass model.
\end{abstract}



%\maketitle

\newcommand{\tw}{t_{\rm w}}

\newcommand{\taup}{\tau_{\rm p}}
\newcommand{\tauh}{\tau_{\rm h}}

\newcommand{\teff}{t_{\rm eff}}

\newcommand{\Lo}{L_{0}}

\newcommand{\Lovlp}{\xi(\delta)}
\newcommand{\LovlpT}{\xi(\Delta T)}

\newcommand{\eq}[1]{Eq.~(\ref{#1})}
\newcommand{\kb}{k_{\rm B}}

\newcommand \be {\begin{equation}}
\newcommand \ee {\end{equation}}

\section{Introduction}

A class of scaling theories \cite{bramoo87,fishus88eq,fishus91} 
for randomly frustrated glassy systems has pointed out a striking fragility
of their free-energy landscapes. While they  realize some glassy order
within a given environment specified for instance by temperature,
even an infinitesimal change of the latter lead to radical 
reformation of the free-energy landscape
to a globally uncorrelated new one. 
Such non-perturbative, global 
shuffling of the free-energy landscape with infinitesimal changes 
of control parameters are called as {\it chaos effects}.
Indeed theoretical studies of some microscopic 
models including studies on Edwards-Anderson (EA) Ising spin-glass models by 
Migdal-Kadanoff renormzalization group (MKRG) method
\cite{banbra87,neyhil93,abm02} and mean-field theory \cite{rizcri2003}
(and references there in) and directed polymers in random media
(DPRM) \cite{salyos2002} have partially or almost 
fully confirmed such striking effects. Further works may clarify
to what extent these unusual phenomena are universal.

A natural interest is to see how slowly relaxing
or aging glassy systems will react to such tortuous perturbations 
\cite{jonetal98,jonyosnor2002,jonetal2003,yoslembou2001,sheyosmaa,sasmar2003}. 
While systems like simple phase separating systems 
would either keep aging accumulatively (domain growth) or 
stop aging under {\it external} driving forces (e.g. stirring oil+ vinegar)
\cite{kur97}, spin-glasses exhibit rejuvenation-memory 
effects \cite{jonetal98} which are far more puzzling and richer.
In \cite{yoslembou2001} a minimal description for such a dynamics
was obtained for the case of Ising spin-glasses  in terms of 
{\it ghost domains}, which  is a direct extension of the concept of 
the standard scaling theory for domain growth \cite{Bray94} 
in isothermal aging. 
In contrast to isothermal aging, the amplitudes of the 
order parameters or bias within domains become  dynamical variables 
which play a central role: they act as {\it internal} driving forces which 
perturb the trajectory of the domain growth itself.
As the result a concrete mechanism of imprinting/retrieving 
multiple memory under the tortuous chaos effects was found.
Recently the  MKRG method was applied to the dynamics 
of the EA model subjected to chaos effects and such a mechanism
was demonstrated explicitly. \cite{sheyosmaa}

In the present paper we revise the ghost domain scenario based on the
theory by Bray and Kisner \cite{BK92} on the growth 
of the bias during domain growth dynamics. 
We consider a simplest one-step ``perturbation-healing'' protocol.
An example is the one-step temperature-cycling protocol 
\cite{lefetal92,andetal93} first used in spin-glasses. 
It proceeds as follows. 
\begin{enumerate}
\item[(1)]  {\it Initial aging stage}.
First a spin-glass is equilibrated 
at a high enough temperature above the glass transition temperature
$T_{\rm g}$. Then at time $t=0$ the temperature 
is quenched down to a temperature say $T_{A}$ below $T_{\rm g}$ 
where the system is aged for some time $\tw$. This stage is just the
same as usual isothermal aging. 
\item[(2)]  {\it Perturbation stage}.  The 
temperature is changed to $T_{B}=T_{A}+\Delta T$ (with $\Delta T$ 
being either positive or negative) where the system is aged for 
some time $\taup$. Strong restart of aging or {\it rejuvenation} 
is observed, for instance, by measuring
the AC magnetic susceptibility in the spin-glasses and
ceramic superconductors \cite{papetal99,garetal2003}.
Other glassy systems such as super cooled liquids \cite{lehnag98}, 
polymer glasses \cite{belcillar2000} exhibit no or much weaker rejuvenations. 
It may suggest absence of chaos effects in some classes of 
glassy systems. One should also keep in mind
that large enough length/time  scales compared with the overlap length 
(See \eq{eq-ovlp}) must be explored to see chaos effects. 
Failures of some experiments and simulations to
detect rejuvenations may be related to this difficulty.
\item[(3)]   {\it Healing stage}. Finally the temperature is put back to
$T_{A}$. In spin-glasses strong restart of aging or rejuvenation is
observed again \cite{andetal93,sasetal2002,jonetal2003}. 
After some {\it recovery time} say $\tau_{\rm rec}$ this restarted process
disappears and the rest of the relaxation becomes
a continuation of the initial aging stage, which 
is called the {\it memory effect}. We closely discuss
the two stage processes in the healing stage based on the 
ghost domain scenario.
\end{enumerate}
Many systems ``heal'' by waiting some  recovery time 
$\tau_{\rm rec}$ after being exposed to a perturbation for 
a certain time $\taup$.  Simple minded ``length scale(s)''
(or some equivalent ``energy-barrier'') arguments 
which neglect the internal driving due to the remanent bias may lead to
two contradictory possibilities:  A) healing is {\it impossible} 
after such strong perturbations due to chaos effects 
or that B) healing is {\it somehow
possible} and the recovery time $\tau_{\rm rec}$ 
is just identical to the time scale at which the length scale 
$L(\tau_{\rm rec})$ (energy barrier) explored after switching 
off the perturbation becomes as large as the length scale $L(\taup)$ 
(energy barrier) explored during the perturbation. 
Furthermore one could argue the ``effective age'' of the system 
imprinted in the system would be largely modified
once $L(\taup)$ becomes larger than the length (energy) 
scale corresponding to the age. 
Somewhat surprisingly we find that all these intuitions fail in general 
for the perturbations operated in the strongly perturbed regime 
of the chaos effect. 
In the present paper we also consider dynamics operated in 
the weakly perturbed regime of the chaos effect. 
This allows us to take into account  effects of slowness
of the switching on/off  perturbations in realistic circumstances.

After introducing the spin-glass model in the next section,
the definition of ghost domains is summarized  and
the scaling theory by Bray and Kisner
is briefly reviewed in section \ref{sec-ghost}.
In sections \ref{sec-cycle1} and \ref{sec-cycle2} 
the revised ghost domain scenario is introduced focusing
on the simple one-step perturbation-healing protocol mentioned above 
and the scenario is examined numerically on
the 4 dimensional EA Ising spin-glass model.
In section \ref{sec-ren-heat-cool} we propose
a simple way to take into account the effects of slow switchings such as
heating/cooling rate effects.
The conclusion of the paper is presented in the last section.

\section{Model}

Specifically we consider the Edwards-Anderson (EA) 
Ising spin-glass models described by a Hamiltonian
\be
H=-\sum_{i,j}J_{ij} S_{i}S_{j}
\label{eq-hamiltonian}
\ee
where $S_{i}$ is an Ising spin at site $i$ located at $\vec{r}_{i}$
on a $d$ dimensional lattice with $N$ lattice sites and $J_{ij}$ 
is a random interaction bond which takes $+J$ and $-J$ 
randomly for each nearest neighbor 
pair $(i,j)$.  Here $J > 0$ is the unit of energy scale.
For convenience we denote the scaled thermal energy $\kb T/J$ 
as temperature $T$ in the following. Here $\kb$ is the Boltzmann's constant.
We consider two kinds of perturbations, 
(1){\it temperature changes}  $T \to T+\Delta T$; 
(2) {\it  bond changes}  ${\cal J} \to {\cal J}'$.  A new set of  bonds
${\cal J}'=\{J'_{ij}\}$ is created from the original one 
${\cal J}=\{J_{ij}\}$  as follows.
For each pair $(i,j)$ we choose $J_{ij}'=-J_{ij}$  randomly with  
probability $p$ and $J_{ij}'=J_{ij}$  with  probability $1-p$.

In the numerical simulation presented in section \ref{sec-cycle2} 
we use the $d=4$ model on the hyper-cubic lattice and 
the single spin flip heat-bath Monte Calro method. In simulations
we limit ourselves to bond changes since computational power is
too limited to study temperature changes  efficiently.

\section{Ghost domains}
\label{sec-ghost}

Let us introduce basic ingredients of the
ghost domain scenario to prepare for
the discussion of the simple one-step perturbation-healing protocol 
(e.~g.~the one-step temperature-cycling experiments)
in the next two sections.
For simplicity we assume that an equilibrium states $\Gamma^{T,{\cal J}}$ 
of a spin-glass a system with a set of random interaction bonds ${\cal J}$
at temperature $T$ below the spin-glass transition temperature $T_{\rm g}$ 
is given by its {\it typical} spin configuration.  
Such a configuration at site $i$
may be described
as $\sqrt{q_{\rm EA}}(T)\sigma^{T,{\cal J}}_{i}$ where $\sigma^{T,{\cal J}}_{i}$ is 
an Ising variable and $q_{\rm EA}(T)$ is the Edwards-Anderson (EA) order parameter
which takes into account the effects of thermal fluctuations.
Furthermore we assume the only possible other
phase at same environment $(T,{\cal J})$ is  $\bar{\Gamma}^{T,{\cal J}}$ 
whose configuration is given by 
$-\sqrt{q_{\rm EA}}(T)\sigma^{T,{\cal J}}_{i}$. However
extensions to the cases that more phases exist for a given environment 
may be considered as well. 

\subsection{Weakly and strongly perturbed regimes of chaos effects}
\label{subsec-weak}

The chaos effects become stronger at larger length scales. 
Since the distinction between the weakly and strongly perturbed 
regimes are important in the following here we summarize the 
picture on the crossover between the two regimes given in
\cite{salyos2002,jonyosnor2003,sheyosmaa}.

Let us consider a generic perturbation which may induce a
droplet excitation of size $L$ with respect to the ``ground state'' 
$\{\sigma_{i}^{T,{\cal J}}\}$. The excited state has a certain free-energy gap 
$F_{L} (>0)$ with respect to the ground state. 
Suppose that we have a  perturbation such that the excited
state obtains a {\it gain} of the free-energy of order
$
\Delta U_{L}/J=\delta (L/\Lo)^{a}.
$
Here $\Lo$ is a microscopic unit length scale.
Then a droplet excitation will be induced if $\Delta U_{L}$
turns out to be greater than the free-energy gap $F_L$. 
The free-energy gap is expected to have
a broad distribution characterized by a distribution function
$\rho_{L}(F_L)$ with the scaling form \cite{bramoo87,fishus88eq},
$
\rho_{L}(F_L)d F_L = \tilde{\rho} 
(F_L/J(L/\Lo)^{\theta})
d F_L/J(L/\Lo)^{\theta}
$
where $J(L/\Lo)^{\theta}$ is the typical free-energy gap with
$\theta (>0)$ being the stiffness exponent .
Using these properties
the probability $p_L(\delta)$ that a perturbation of strength $\delta$
induces a droplet excitation of size $L$ is found as
\be
p_L(\delta) \sim  \int_{0}^{\Delta U_{L}} d F_L \rho(F_L )
= \int_{0}^{(L/\Lovlp)^{\zeta}} d y \tilde{\rho}(y)
\label{eq-prob-event}
\ee
where 
\be
\Lovlp=\Lo \delta^{-1/\zeta}
\label{eq-ovlp}
\ee
is the characteristic crossover length, called overlap length, 
beyond which $p_L(\delta)$ becomes $O(1)$. The exponent $\zeta$, the so
called chaos exponent, is given by
$
\zeta=a - \theta.
$
One can see that if $\zeta >0$ ($a > \theta$) the 
probability $p_L(\delta )$ 
continuously increases  with increasing $L/\Lovlp$. 
In the following we distinguish between 
the {\it strongly perturbed regime} $L/\Lovlp > 1$ and
the {\it weakly perturbed regime} $L/\Lovlp < 1$.

In the {\em strongly perturbed regime} $L/\Lovlp > 1$, 
the original ground state
$\{\sigma_{i}^{T,{\cal J}}\}$ is completely unstable with respect to the
droplet excitations, i.e. a new equilibrium state must form.  
The term {\it chaos} \cite{bramoo87,fishus88eq,fishus88noneq,fishus91} 
properly describes
the fact that a strongly perturbed regime eventually emerge even for 
arbitrary small $\delta \ll 1$ at sufficiently large length scales.
However, chaos does not set in
abruptly at the overlap length $\Lovlp$ but, 
in the {\em weakly perturbed regime} $L/\Lovlp < 1$ 
chaos like droplet excitations already occur at length scales smaller than $\Lovlp$
with non-zero probability $p_{L}(\delta)$
\cite{salyos2002,jonyosnor2003,sheyosmaa}.

In the case of temperature shifts of strength $\Delta T$ 
the possible free-energy gain of a droplet excitation 
of size $L$ is the entropy gain ($\times  \Delta T$) 
which is expected to scales as $\Delta U_{L}/J \sim \Delta T(L/\Lo)^{d_{s}/2}$
where $d_s$ is the surface fractal dimension of droplet excitations.
(See \cite{abm02} for a detailed discussion)
In the case of bond perturbations, the random gain of energy of a 
droplet excitation happen at around its surface so that 
$\Delta U_{L}/J \sim p  (L/\Lo)^{d_{s}/2}$.
Thus temperature and bond perturbations
should lead to a chaos effect of the same universality class with
$\zeta=d_{s}/2-\theta$.

\subsection{Definition of ghost domains}

Let us consider a generic protocol such the working environment is
changed from time to time among a set of target environments
$\{A,B,\ldots\}$ which consists of different temperatures
$\{T_{A},T_{B},\ldots\}$ (all below $T_{g}$ )
and/or different bonds $\{{\cal J}_{A},{\cal J}_{B},\ldots\}$
whose equilibrium states are represented by 
${\  \sqrt{q_{\rm EA}(T_{A})}\sigma^{A}_{i},
\sqrt{q_{\rm EA}(T_{B})}\sigma^{B}_{i},\ldots}$. 

Suppose that the system is now evolving in a 
certain working environment, say 
$W=(T_{W},{\cal J}_{W})$ at a certain time $t$.
Short time averages may be took to average out 
short time thermal fluctuations. Then the temporal 
spin configuration can be represented as 
$\sqrt{q_{\rm EA}(T_{W})}s_{i}(t)$ where $s_{i}(t)$
takes Ising values.
It can be projected onto the equilibrium 
states of {\it any} environment  $R \in \{A,B,\ldots \}$ as
\be
\tilde{s}^{R}_{i}(t)=\sigma_{i}^{R}s_{i}(t).
\ee
Then the projected image $\tilde{s}^{R}_{i}(t)$ is described 
in a coarse-grained way by the following two features.
\begin{enumerate} 
\item the {\it domain wall configuration}: configuration of 
the spatial pattern of the {\it sign} of the projection
 $\tilde{s}^{R}_{i}(t)$.

\item the {\it order parameter}: the amplitude of the projection
$\rho^{R}(t)=|[\tilde{s}^{R}_{i}(t)]_{\rm domain}|$  
where $[\ldots]_{\rm domain}$ denote the spatial average within a ghost domain.
\end{enumerate}
It is useful to consider decomposition of a 
ghost domain $\Gamma^{R}$ ($\bar{\Gamma}^{R}$) 
into ``patches'',
\begin{itemize}
\item The strength of the bias has the full amplitude $1$ within a patch.
\item The ``signs $+/-$'' of the bias is however different on different
      patches-the majority has the same sign as that of the ghost domain
      to which they belong to. Minorities have the opposite sign.
\end{itemize}
The probability $p_{\rm minor}(t)$ that a patch belongs 
to the {\it minority phase} $\bar{\Gamma}^{R}$ ($\Gamma^{R}$) 
in a ghost domain of $\Gamma^{R}$ ($\bar{\Gamma}^{R}$)
is related to the strength of the bias  $\rho^{R}(t)$ as
\be
p_{\rm minor}(t)=(1-\rho^{R}(t))/2.
\label{eq-p-minority}
\ee

If one chooses $R=W$, a ghost domain reduces to an 
ordinary domain which is enough 
in isothermal aging where the order parameter is a constant.
In the cycling protocols, {\it minimal} description is to keep track
of projections on to the equilibrium states of
{\it all} target environments. We call such projections 
as {\it ghost domains}. Very important point is that  not 
only (i) the domain wall structures 
 but also (ii) the amplitude of the order parameter within the domains are 
dynamical variables. 

\subsection{Physical observable}
\label{subsec-physobs}

Let us note here that basic quantities measured  
in experiments and simulations are essentially {\it gauge invariant}
(or {\it ghost invariant}), 
i.~e.~they do not depend on specific choice of projections.
An important example is the spin auto-correlation function 
\be
C(t,t')=(1/N) \sum_{i}<S_{i}(t)S_{i}(t')>
\label{eq-def-ac}
\ee
where $N$ is the number spins and $<\ldots>$ represents taking an averages
over different trajectories. The auto-correlation function
can be re-expressed in terms of any projection field
$\tilde{s}^{R}_{i}(t)$ as $C(t,t')=q_{\rm EA}(T_{R})(1/N) \sum_{i}<
\tilde{s}^{R}_{i}(t)\tilde{s}^{R}_{i}(t')>$
because $\sigma^{R}_{i}=\pm 1$. Thus it
does not depend explicitly on the specific choice of projections, 
i.e. gauge invariant except for the prepfactor. 
The auto-correlation function can be measured experimentally
by monitoring spontaneous thermal fluctuations of the
magnetization $M(t)$ \cite{HO02} because the leading $O(N)$ part
of the magnetic auto-correlation function is $NC(t,t')$ in spin-glasses
with no ferromagnetic or anti-ferromagnetic bias in the distribution of
the bonds. In experiments linear magnetic susceptibilities
to uniform external magnetic field are often measured.
By the same token as above it can be seen that  the linear magnetic 
response functions (per spin) 
of spin-glasses to a uniform external field  $h(t')$,
$R(t,t')=(1/N)\partial <M(t)>/\partial h(t')$ is essentially gauge-invariant. 

\subsection{Basic dynamics at a working environment}

Suppose that the system is temporally evolving at a certain working 
environment $W$ with temperature $T=T_{W}$ with a certain set of bonds ${\cal J}_{W}$. 
Here we summarize basic
properties of the dynamics of the (ghost) domains of $\Gamma^{W}/\bar{\Gamma}^{W}$.

At coarse-grained mesoscopic level, the relaxational dynamics is 
considered as a thermally activated process of a droplet like
excitation. The energy barrier associated with a droplet of size $L$
scales as $E_{\rm b} \sim \Delta (T)(L/\Lo)^{\psi}$ with $\psi > 0$.
Thus at a given logarithmic time scale $\log(t/\tau_{0}(T))$ 
a droplet as large as  
\begin{equation}
L_{T}(t)
\sim L_{0}
[ (\kb T/\Delta(T))
\ln (t/\tau_{0}(T))]^{1/\psi}
\label{eq-growth} 
\end{equation}
can be thermally activated \cite{fishus88noneq}. In the above formula
the effects of critical fluctuations can be took into account 
in a renormalized way in the characteristic energy scale  $\Delta(T)$ 
for the free-energy barrier and the characteristic time 
scale $\tau_{0}(T)$.

Suppose that the projection of the initial spin configuration
onto the equilibrium state $\Gamma^{W}$ at time $t=0$ is strongly 
disordered such that its spatial correlation function decays
rapidly beyond some correlation length $\xi_{\rm ini}$,
\begin{eqnarray}
&& [(\tilde{s}^{W}_{i}(0)-\rho^{W}(0))(\tilde{s}^{W}_{j}(0)-\rho^{W}(0))] = F
\left(\frac{|\vec{r}_i-\vec{r}_j)|}{{\xi_{\rm ini}}}\right)\nonumber \\
&& [\tilde{s}^{W}_{i}(0)] =  \rho^{W}(0)
\label{eq-random-ini-bias}
\end{eqnarray}
Here $[\ldots]$ denotes the average over space and $F(x)$ 
is a certain rapidly decreasing function.
Note that the bias $\rho^{W}(0)$ is made {\it homogeneous} within the system.

{\bf Domain growth without bias-}
If the initial bias is absent $\rho^{W}(0)=0$,
the global $Z_{2}$ symmetry of the system is not broken and 
the domain growth (aging) never stops.
The mean separation between the domain walls at time $t$
is $L_{T}(t)$ given in \eq{eq-growth} \cite{fishus88noneq}. 
In such a {\it critical quench} the spatial correlation function
\be
C^{W}(r,t,t')=[<\tilde{s}^{W}_{i}(t)\tilde{s}^{W}_{j}(t')>]_{r=|\vec{r}_{i}-\vec{r}_{j}|}
\label{eq-def-crt}
\ee
exhibits universal scaling properties \cite{Bray94}.
In the so called aging regime $L_{T}(t) > L_{T}(t')$ it scales as
\be
C^{W}_{0}(r,t,t') \sim 
\left(\frac{L_{T}(t)}{L_{T}(t')}\right)^{-\bar{\lambda}}
h \left(\frac{r}{L_{T}(t')}\right) \qquad L_{T}(t) > L_{T}(t').
\label{eq-scaling-crt}
\ee
The subscript ``$0$'' is meant to emphasize that the spin configuration
is random at $t=0$ with respect to the target equilibrium state.
Here  $h(x)$ is a decreasing function with $h(0)=1$.
The exponent $\bar{\lambda}$ is a non-equilibrium dynamical 
exponent introduced by Fisher-Huse in \cite{fishus88noneq}. 
(Note that in some literatures e.g. Refs \cite{fishus88noneq} 
(and also \cite{yoslembou2001,yoshuktak2002} )
$\bar{\lambda}$ is denoted as $\lambda$. In the present paper 
we follow \cite{BK92,Bray94} and use
$\bar{\lambda}$ for the decay of the correlation function and $\lambda$
for the growth of bias discussed below (See \eq{eq-bias-growth}).)

A special case of much interest is the auto-correlation function
($r=0$) which is just the spin auto-correlation function $C(t,t')$
defined in \eq{eq-def-ac} which is a gauge invariant quantity.
It generically follows a scaling
of the form $C_{0}(t,t')=C_{0}(L_{T}(t)/L_{T}(t'))$. The scaling function
$C_{0}(x)$ remains at $1$ 
in the quasi-equilibrium regime $x<1$. 
In the aging regime $x >1$, it approaches $0$ asymptotically as
$C_{0}(x) \sim x^{-\bar{\lambda}}$. In Ising spin glasses
\be
d/2 \leq  \bar{\lambda} <  d
\label{eq-barlambda-isingsg}
\ee
is proposed \cite{fishus88noneq}.
This very slow relaxation is in sharp contrast to the exponential
decay in the paramagnetic phase
$C_{0}(t,t')  \propto \exp(-|t-t'|/\tau_{\rm eq}(T))$
where $\tau_{\rm eq}(T)$ is the correlation time in the paramagnetic phase.

{\bf Domain growth with bias-}
Even if the initial bias $\rho^{W}(0)$ is small, the $Z_{2}$ symmetry is
explicitly broken if it is non-zero. One expects that the strength of
the symmetry breaking will increase with time and eventually terminates 
the aging just as if {\it external} symmetry breaking field is applied.
This problem was considered theoretically first
by Bray and Kisner \cite{BK92} . They noticed that the non-zero {\it homogeneous}
bias grows with time $t$ as
as
\be
\rho^{W}(t) \sim  \rho^{W}(0) \left(\frac{L_{T}(t)}{\xi_{\rm ini}}\right)^{\lambda}
\label{eq-bias-growth}
\ee
and the dynamical exponent $\lambda$ is related to  $\bar{\lambda}$
as
\be
\bar{\lambda}+\lambda=d
\label{eq-lambda-relation}
\ee
Here let us summarize the derivation \cite{BK92} within our context.  
First one can see that the bias is nothing but 
the $k=0$ component of the Fourier 
transform of $\tilde{s}_{i}^{W}$. Then one assumes that
the amplitude of the $k=0$ component at time $t$ can be
computed as a linear-response to the change of its initial value.
Second assuming the Gaussian characteristics of the random initial
condition \eq{eq-random-ini-bias} one finds the linear response 
function is the same as the $k=0$ component of 
the spin correlation function \eq{eq-def-crt} up to
some proportionality constant $c$.  As the result one 
obtains
\be
\rho^{W}(t) = c  \rho^{W}(0) C_{k=0}(t,0)
\ee
Then using \eq{eq-def-crt} one finds \eq{eq-bias-growth} and \eq{eq-lambda-relation}.
Combining the scaling relation \eq{eq-lambda-relation} and
the inequality \eq{eq-barlambda-isingsg} one finds
\be
\bar{\lambda}/\lambda \geq  1.
\label{eq-ratio-lambda}
\ee
As we discuss below this inequality suggests healing of spin-glasses
after chaotic perturbations takes an enormously long time. \cite{foot1}

\section{A cycle on a globally symmetry broken state}
\label{sec-cycle1}

\begin{figure}[b]
\begin{center}
\includegraphics[width=0.9\columnwidth]{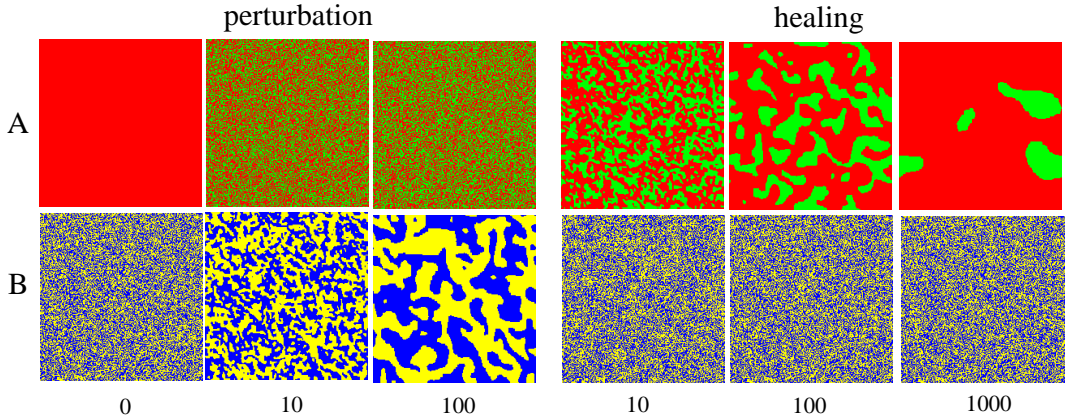}
\end{center}
\caption{Evolution of ghost domains in a simple ``perturbation-healing'' 
protocol on a globally symmetry broken state.
This is a demonstration using a heat-bath Monte Calro 
simulation of 2-dim Ising Mattis model ($N=400 \times 400$)
in which the interaction bonds in \eq{eq-hamiltonian} are 
given as $J_{ij}=J \sigma_{i}\sigma_{j}$
where $\sigma_{i}$ is a  random Ising (gauge) variables given at each site.
One immediately finds the equilibrium state (ground state) for each ${\cal J}$ 
is simply given by the set $\{\sigma_i\}$.
The initial spin configuration is chosen to be identical to 
a random ground state $\sigma^{A}$.
{\bf Perturbation}: For time $\tau_{p}=100$ (MCS) the system 
is strongly perturbed by using 
the Hamiltonian of a different ground state $\sigma^{B}$ 
which is completely uncorrelated with $\sigma^{A}$:
$\Lovlp=1$  in the unit lattice. 
{\bf Healing}: Then the Hamiltonian is put back to the original one
which is used for additional $1000$ (MCS). In the present examples 
temperature is set to  $T=2.0$.
The different colors (greycales) represent the sign $+$ and $-$ of 
the projections on to the ground states.
The 3 columns on the left sides
are snapshots at time $0, 10, 100$ (MCS) during the perturbation.
Domains of $\Gamma^{B}/\bar{\Gamma}^{B}$ grow while the bias $\rho^{A}$ decreases.
In this example $\rho^{A}$ has become $0.03$ which is too small to
 distinguish by eye. In the 3 columns on
the right  are snapshots at time $10, 100, 1000$ (MCS) 
during the healing.
Here domains of $\Gamma^{A}/\bar{\Gamma}^{A}$ grow but 
the minority phase $\bar{\Gamma}^{A}$ slowly disappears and $\rho^{A}$ increases. 
The recovery of the $\rho^{A}$ turns out to be much longer time than
$\tau_{p}$. In this model the growth law \eq{eq-growth} should
be replaced by $L(t)/\Lo \sim \sqrt{t/t_{0}}$  since it is 
equivalent to the ferromagnetic Ising model \cite{Bray94}.
We found numerically $\lambda \sim 0.8$ and $\bar{\lambda}\sim
1.2$ in this model being consistent with \eq{eq-lambda-relation}}
\label{fig-ghost}
\end{figure}

Let us now begin with the perturbation-healing protocol 
by considering an idealized limit.
This corresponds for example to the one-step temperature-cycling experiments
mentioned in the introduction 
$T_{A} \to T_{B} \to T_{A}$ but with the initial aging
done for an extremely long time  $\tw=\infty$:
the spin configuration is globally equilibrated with respect to
an equilibrium state $\Gamma^{A}$ at the beginning. 
Then {\it perturbation} is performed - 
change temperature or bond and let the system evolve 
for a certain time $\tau_{p}$ so that domains of 
$\Gamma^{B}/\bar{\Gamma}^{B}$ grow.
Lastly {\it healing} is performed - switch off the perturbation and let 
the system evolve after wards for some time  $\tauh$. 
Here $A$ and $B$ would stand for (1) $T_{A}$ and $T_{B}=T_{A}+\Delta T$
in the case of temperature-cycling or (2)
${\cal J}_{A}$ and ${\cal J}_{B}$ (which is created 
by randomly changing the sign of a fraction $p$ of the bonds
in the original set ${\cal J}_{A}$) in the case of bond-cycling.
For simplicity we assume the switchings are instantaneous. 
The effects of slow switching times, e.g.
finite heating/cooling rates,  will be discussed later in section
\ref{sec-ren-heat-cool}.

{\bf Strongly perturbed regime-}If the duration of the perturbation
$\tau_{p}$ is long enough such that $L_{B}(\tau_{p}) > \Lovlp$, 
where $\Lovlp$ is the overlap length given in \eq{eq-ovlp} and $\delta$
can be either $\Delta T$ or $p$,
the strongly perturbed regime of the chaos effects 
(see section \ref{subsec-weak}) should come into play.
For simplicity here we neglect dynamics at short time scales
which belong to the weakly perturbed regime.
An extreme example of $\Lovlp=1$
is shown in Fig \ref{fig-ghost} using a Monte Calro simulation of 
a 2-dimensional Ising Mattis model.

The initial spin configuration is globally aligned to
$\Gamma^{A}$ so that it is fully biased as $\tilde{s}^{A}_{i}=1$.
But simultaneously $\tilde{s}^{B}_{i}$ is completely disordered (beyond $\Lovlp$) 
with no bias $[<\tilde{s}^{A}_{i}>]=0$.
Thus during the perturbation stage, the domains 
of both $\Gamma^{B}/\bar{\Gamma}^{B}$ grow competing 
with each other just as the usual domain growth. Their typical size
becomes $L_{B}(\taup)$.
As can be seen in the Figure \ref{fig-ghost}, this amounts to reduction of the 
bias (or staggard magnetization) with respect to $\Gamma^{A}$.
The remanent bias $\rho_{\rm rem}^{A}$ becomes
\begin{eqnarray}
 \rho_{\rm rem}^{A}(\taup) &=&(1/N)\sum_{i}\tilde{s}^{A}_{i}(\taup)
=(1/N)\sum_{i}\tilde{s}^{A}_{i}(0)\tilde{s}^{A}_{i}(\taup)
=(1/N)\sum_{i}\tilde{s}^{B}_{i}(0)\tilde{s}^{B}_{i}(\taup)
 \nonumber \\
& =& C_{0}(\tau_{p},0)  \sim  (L_{B}(\taup)/\Lovlp)^{-\bar{\lambda}}.
\label{eq-rho-rem}
\end{eqnarray}
Here we used the properties of the initial condition and the gauge
invariance of the auto-correlation function. The subscript ``0''
is used in the last equation 
because the initial condition is completely random with respect
to the relaxational dynamics which is {\it at work} during the perturbation.
Here it can be seen clearly that the strong perturbation due to the chaos effect
is {\it not} equivalent to put a system to a paramagnetic phase. In the
latter case one would find exponential decay of the bias 
$\rho_{\rm rem}^{A}(\taup) \sim \exp(-\taup/\tau_{\rm eq}(T_{B}))$ where
$\tau_{\rm eq}(T_{B})$ is the relaxation time in the paramagnetic phase.
Thus chaos effect do {\it not} amount pushing the system to the
disordered phase contrary to what might have been suspected.

During the healing stage, the domains of both $\Gamma^{A}/\bar{\Gamma}^{A}$ 
grow competing with each other starting from the minimum length scale
$\Lovlp$. Their typical size becomes $L_{A}(\tauh)$.
However this is a domain growth with biased initial condition
as \eq{eq-random-ini-bias}.
From \eq{eq-bias-growth} the strength of the global bias is found to grows as
\be
\rho^{A}_{\rm rec}(\tauh)=\rho_{\rm rem}^{A} (L_{A}(\tauh)/\Lovlp)^{\lambda}
\label{eq-rho-rec}
\ee
where $\rho_{\rm rem}^{A}$ is the remanent bias at the end of
the perturbation stage (or  the beginning of the healing stage).
Note that $\rho^{A}_{\rm rec}(\tauh)$ is {\it proportional to}
the initial remanent bias $\rho_{\rm rem}^{A}$  which is the 
direct consequence of the ``linear-response' of the temporal 
bias with respect to the change of the initial bias 
as noticed by Bray and Kisner. 

The above situations may be rephrased as the following.
During the perturbation stage the system may be decomposed into patches
whose size is given by the overlap length $\Lovlp$. 
The probability $p_{\rm minor}$ that 
a patch belong to the {\it minority phase} $\bar{\Gamma}^{A}$ 
is related to the bias $\rho^{A}$ as given in \eq{eq-p-minority}
so that it increases with $\tau_{p}$ during the perturbation stage.
Next in the healing stage the system may be decomposed into 
patches of size $L_{A}(\tauh)$, which now grows with $\tau_{h}$. 
The probability $p_{\rm minor}$ now decreases because $\rho^{A}$ increases.
Consequently the  minority phase eventually disappears 
and the domain growth stops at the recovery time $\tau_{\rm rec}$.

Combining \eq{eq-rho-rem} and \eq{eq-rho-rec} one finds the recovery
time $\tau^{\rm strong}_{\rm rec}$ of the strength of the bias as
\be
\frac{L_{A}(\tau^{\rm strong}_{\rm rec})}{\Lovlp}=\left(\frac{L_{B}(\taup)}{\Lovlp}
\right)^{\bar{\lambda}/\lambda}
\label{eq-tau-rec}
\ee 
Here the super-script ``strong'' is mean to emphasize that it is
a formula valid in the strongly perturbed regime.
Since $\bar{\lambda}/\lambda \geq  1$ as given in \eq{eq-ratio-lambda},
we conclude that the recovery time can be significantly 
large. \cite{foot1}

The above considerations can be extended to multi-step cycling.
Very counter-intuitive consequences follow due to multiplicative nature of 
the effect of multiple perturbations \cite{yoslembou2001}.
For example another perturbation stage to grow $\Gamma^{C}/\bar{\Gamma}^{C}$
for some time $\tau'_{\rm p}$ can be added before the healing stage in the 
perturbation-healing protocol  discussed above.
Here $\Gamma^{C}$ is assumed to be decorrelated with respect to
both $\Gamma^{A}$ and $\Gamma^{B}$ beyond the overlap length $\Lovlp$.
Then the recovery time $\tau^{\rm strong}_{\rm rec}$ of 
the order parameter of $A$ becomes,
\be
\frac{L_{T_A}(\tau^{\rm strong}_{\rm rec})}{\Lovlp}
= \left(\frac{L_{B}(\taup)}{\Lovlp}\frac{L_{C}(\tau'_{\rm p})}{\Lovlp}\right)^{\bar{\lambda}/\lambda}
\label{eq-tau-rec-multi}
\ee
which can yield huge recovery time.

{\bf Weakly perturbed regime-} If the duration of the perturbation
$\tau_{p}$ is small such that $L_{B} < \Lovlp$, the effect
of the perturbation should remain mild as explained 
in section \ref{subsec-weak}.
Here the mutual interferences between ghost domains just amount to
induce {\it rare} droplet excitations of various size $L$  with probability
$p_{L}(\delta) \ll 1$ (see \eq{eq-prob-event}) on top of each other.  
They are just independent islands of the minority phase which rarely
overlap with each other. Thus one only needs to keep track of 
switch on/off of such isolated objects during  perturbation 
and healing stages.
This means a naive ``length scale(s)'' argument 
to estimate the recovery time of bias fortunately do not fail in this regime.

More precisely the result of \cite{sheyosmaa} implies 
the remanent bias decreases due to the increase of rare islands of the minority phase as
\be
\rho^{A}_{\rm rem}(\taup)=1-c p_{L_{B}(\taup)}(\delta) +O(p^{2}) \simeq 1-c
(L_{B}(\taup)/\Lovlp)^{\zeta}
\label{eq-rho-rem-weak}
\ee
in the perturbation stage and increases as 
\be
\rho^{A}_{\rm rec}(\tauh)=\rho_{\rm rem}(\taup)+c (L_{A}(\tauh)/\Lovlp)^{\zeta}
\label{eq-rho-rec-weak}
\ee
by removing islands one by one 
in the healing stage.  Here $c$ is some numerical constant.
We have neglected higher order terms of $O(p^{2})$.
In the MKRG analysis \cite{sheyosmaa}, it has been shown that
both \eq{eq-rho-rem-weak} and \eq{eq-rho-rem} are limiting behaviors
of a universal scaling function of of $L_{B}/\Lovlp$.
Note that the bias recovers in {\it additive} fashion in \eq{eq-rho-rec-weak}
which is markedly different from the {\it multiplicative} fashion 
found in the strongly perturbed regime  \eq{eq-rho-rec}. 

One finds the recovery time in the weakly perturbed regime is simply given as
\be
L_{A}(\tau^{\rm weak}_{\rm rec})=L_{B}(\taup) \qquad \mbox{or} \qquad
\tau^{\rm weak}_{\rm rec}/\tau_{0}(T_{A}) =
(t/\tau_{0}(T_{B}))^{(\Delta(T_{A})/\Delta(T_{B}))(T_{B}/T_{A})}
\label{eq-tau-rec-weak}
\ee 
Here the super-script ``weak'' is mean to emphasize that it is
a formula valid only in the weakly perturbed regime. The 2nd
equation holds in the case of activated dynamics \eq{eq-growth}
which simplifies further at low enough temperature 
as $\tau^{\rm weak}_{\rm rec}/\tau_{0}=(t/\tau_{0})^{(T_{B}/T_{A})}$
where temperature dependence of the unit time $\tau_{0}(T)$ and the 
energy scale $\Delta(T)$ can be neglected. There one only
needs to know the microscopic time scale $\tau_{0}$, which is known to
be around $10^{-12}-10^{-13}$ (sec) in real spin-glass materials, to estimate
the recovery time $\tau^{\rm weak}_{\rm rec}$.

Moreover it is easy to see that non-overlapping islands
of the minority phase cannot cause any non-trivial 
effect of multiple perturbations (see
\eq{eq-tau-rec-multi}). This point becomes important when we consider
the effects of slow switching, e.~g. heating/cooling rate effects
in section \ref{sec-ren-heat-cool}.

Other non-chaotic, {\it mild} effects of perturbations can be
considered in similar ways. For instance change of
thermally active droplets can be took into account by changing
$p_{L}(\delta)$ above by $\Delta T/(L/\Lo)^{\-\theta}$. The latter
amounts to an even weaker effect at larger time scales 
but may be dominant at short time scales.

\section{Parallel evolution of domain sizes and order parameter}
\label{sec-cycle2}

Let us complete the scenario for the one-step perturbation-healing protocol
by now allowing the waiting time $\tw$ in the initial aging to be 
{\it finite}.
Suppose  the system is completely disordered with respect
to both $\Gamma^{A}$ and $\Gamma^{B}$ at the beginning 
and aged for some waiting time $\tw$
at $A$. Then instead of an infinitely large domain of $\Gamma^{A}$,
there will be domains of $\Gamma^{A}$ and $\bar{\Gamma}^{A}$ of
size $L_{A}(\tw)$. In the following we consider 
the perturbation-healing protocol exerted onto this initial state.

The time evolution of the system in
a cycling protocol can be concisely described by the time evolution 
of the ghost domains of $\Gamma^{A}/\bar{\Gamma}^{A}$ 
and $\Gamma^{B}/\bar{\Gamma}^{B}$. \cite{yoslembou2001}
The situation may be visualized  again simply by patches.
A ghost domain $\Gamma^{A}$ of size $L_{A}(\tw)$ may be 
decomposed into patches of size $\Lovlp$ during perturbation stage 
and patches of size $L_{A}(\tauh)$ during healing stage.
Within a patch  the bias is always homogeneous 
and has the full amplitude $1$.  But the ``signs'' of the
bias is different on different patches. The probability 
$p_{\rm minor}$ (\eq{eq-p-minority}) that a patch belong to the {\it minority phase} 
$\bar{\Gamma}^{A}$ increases with time as \eq{eq-rho-rem} in the
perturbation stage and decreases as \eq{eq-rho-rec} in the healing stage.
Since the size of the ghost domain itself is finite, it also continues to
grow during the healing stage. Following \cite{yoslembou2001} we call the new 
domain growth under the background bias field during the healing 
as {\it inner-coarsening} and the further growth of the size of the 
ghost domain itself as {\it outer-coarsening}.

The crucial point is that the projection $\tilde{s}^{A}$
keeps the same long wavelength spatial structure
of {\it sign} of the bias as the original domain structure just 
before the perturbation throughout the perturbation-healing stages. 
In the absence of such an explicit mechanism of a sort of symmetry breaking, 
the new  domains grown during the healing 
would have completely wrong 
signs of bias and could lead to total erasure of the memory. (The
scenario (A) mentioned in the introduction.) 
The latter was the main problem in the previous attempt to model 
multiple domains by Koper and Hilhorst \cite{kophil88} and many other
popular ``length scale(s)'' arguments which neglect the role of the internal
driving by the remanent bias.

\subsection{Memory correlation function}

The memory of the ``state'' of the system just before the perturbation
(or the end of the initial aging stagey)
can be directly quantified by the spin auto-correlation function as,
\be
C_{\rm mem}(\tau+\tw,\tw)=C(\tau+\tw,\tw).
\label{eq-c13}
\ee
which we call as {\it memory correlation function}.
The hamming distance $d=1-C_{\rm mem}$ gives a measure of the closeness in the
phase space between the phase points at the two times.
Note that in the limit $\tw \to \infty$ it reduces to the 
global bias $\rho^{A}$ discussed in section \ref{sec-cycle1}.
It is useful to recall that 
the auto-correlation function is gauge invariant so that it is suitable for
experiments/simulations of spin glass systems where one does not 
know {\it a priori} any equilibrium states below $T_{\rm g}$.

\begin{figure}[t]
\begin{center}
\includegraphics[width=0.7\columnwidth]{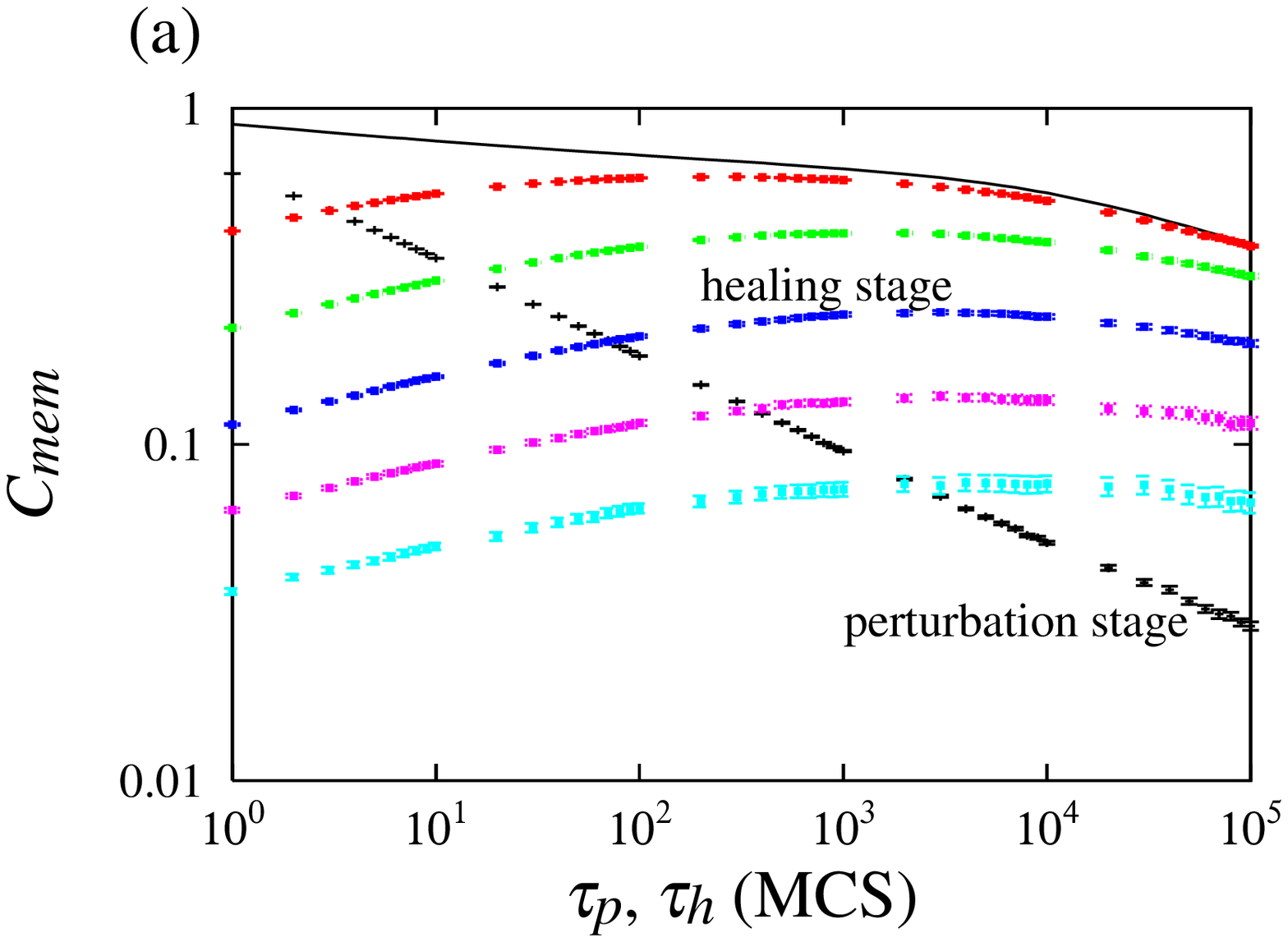}
\includegraphics[width=0.7\columnwidth]{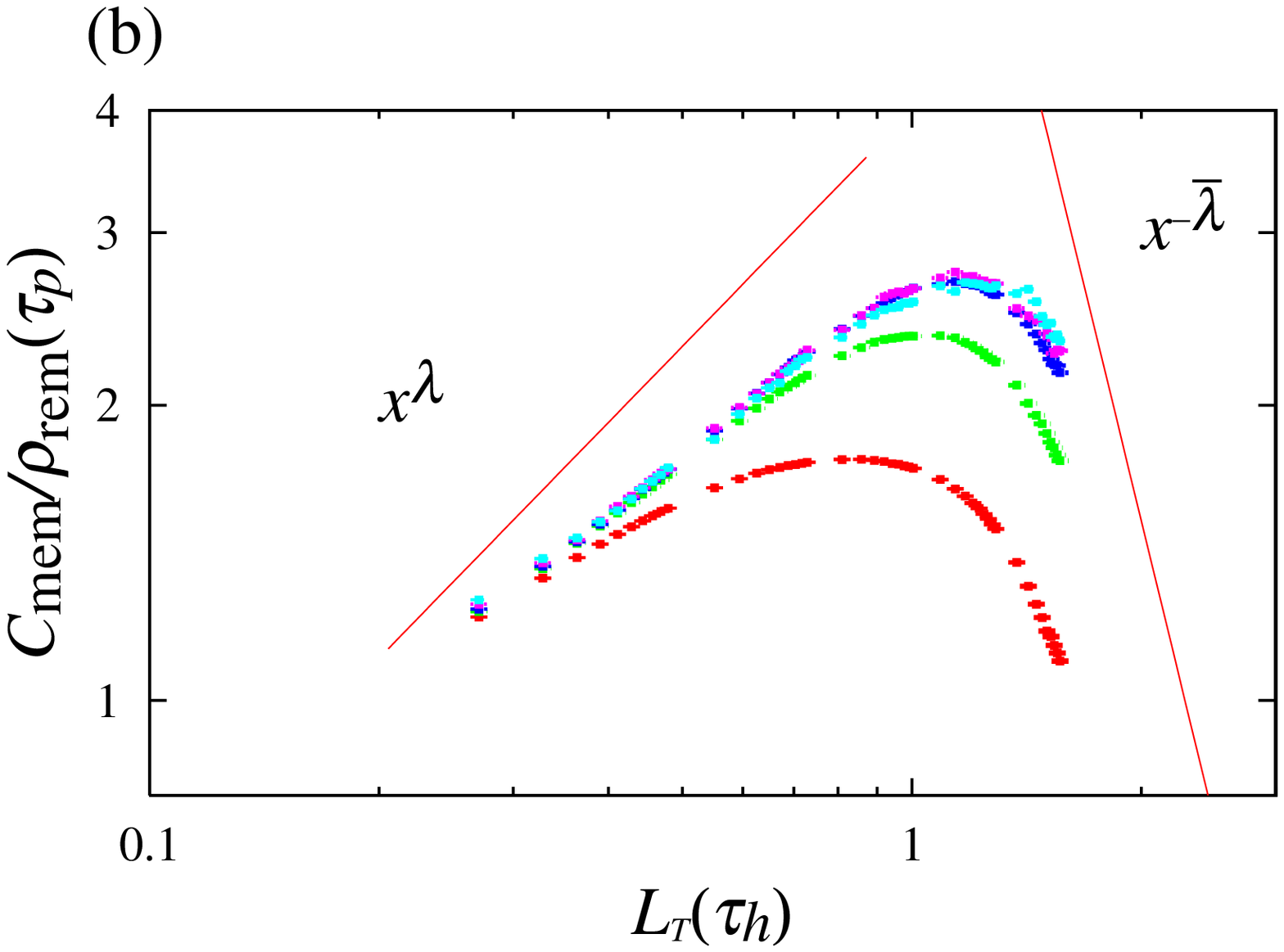}
\end{center}
\caption{Memory correlation function of the 4 dim EA model.
The perturbation is a bond perturbation of strength $p=0.2$.
The temperature is $T=1.2$ ($T_{\rm g} = 2.0$).
The system size is $N=24^{4}$ which is large enough to avoid
finite size effects within the present time window \cite{yoshuktak2002}.
The initial waiting time is fixed as $\tw=10^{4}$ (MCS).
(a) The data points labeled ``perturbation stage'' is 
$C_{\rm mem}(\taup+\tw,\tw)$. Other data points are those 
in the healing  stage $C_{\rm mem}(\tauh+\taup+\tw,\tw)$
after various duration of the perturbation 
$\tau_{\rm p}=10,10^{2},10^{3},10^{4},10^{5}$ (MCS)
from the top to the bottom. Note that the last one is even larger
than the initial waiting time $\tw$.
The solid line is the reference curve of $C_{0}(\tauh+\tw,\tw)$
obtained by a simulation of isothermal aging with $\tw=10^{4}$ (MCS).
(b) The memory correlation functions scaled by the {\it remanent bias}
$\rho_{\rm rem}(\taup)$ are shown. The latter is directly read off
from the data in the perturbation stage  (shown in (a))
as $\rho_{\rm rem}(\taup)=C_{\rm mem}(\tau_{\rm p}+\tw,\tw)$.
Here the dynamical length $L_{T=1.2}(t)$ is used which has been
obtained in a previous study of the same model \cite{hukyostak2000,yoshuktak2002}.
The straight lines are the power laws $x^{\lambda}$ and
 $x^{-{\bar{\lambda}}}$ with $\lambda=0.8$ and $\bar{\lambda}=3.2$.
}
\label{fig-4dea}
\end{figure}

{\bf Strongly perturbed regime-}  Let us first consider
a cycling operated in the strongly perturbed regime.
During the perturbation stage one can easily see
$C_{\rm mem}$ is identical to  $\sqrt{q_{\rm EA}(T_{A})}\sqrt{q_{\rm EA}(T_{B})}
C_{0}(\taup,0)$ where $C_{0}(\taup,0)=\rho^{A}_{\rm rem}(\taup)=(L_{B}(\taup)/\Lovlp)^{-\bar{\lambda}}$
(See \eq{eq-rho-rem})
During the healing stage
the analytical result of a spherical Mattis model 
suggest the following factorization (See Eq. (109) of  \cite{yoslembou2001})
\be
C_{\rm mem}(\tauh+\taup+\tw,\tw)
=q_{\rm EA}(T_{A})\rho^{A}_{\rm rec}(\tauh; \rho^{A}_{\rm rem}) C_{0}(\tauh+\tw,\tw).
\label{eq-c-memory}
\ee
Here the factor $\rho^{A}_{\rm rec}(\tauh;\rho^{A}_{\rm rem})$ 
represents the growth of the bias within the ghost domains by the
{\it inner-coarsening} 
(See \eq{eq-rho-rec}) and  the factor $C_{0}(\tauh+\tw,\tw)$ 
represents the outer-coarsening which is the auto-correlation function 
in isothermal aging (without perturbation $\taup=0$).
Thus the memory correlation function \eq{eq-c-memory} 
behaves non-monotonically with time in the healing stage
and exhibits a peculiar ``memory peak'' because of 
the two competing factors: 
$\rho_{\rm rec}(\tauh)$ increases while $C_{0}(\tauh+\tw,\tw)$
decreases with $\tau_{\rm h}$.

In Figure \ref{fig-4dea} we show the memory auto-correlation function
of the 4 dim EA model after a bond perturbation of strength $p=0.2$.
By an independent numerical study of a bond-shift simulation
done in the same way as in a recent temperature-shift 
experiment \cite{jonyosnor2002,jonyosnor2003}, we checked our time window
lies almost entirely in the strongly perturbed regime with $p=0.2$ 
as reported elsewhere \cite{jonetal2003}. In the scaling plot (b), 
the expected {\it multiplicative} recovery of bias (memory) 
(See \eq{eq-rho-rec}) is demonstrated.

In a previous study of this model \cite{yoshuktak2002}
$\bar{\lambda} \sim 3.0 - 3.5$ was found by analyzing 
the relaxation of thermo remanent magnetization (TRM).
As shown in the scaling plot (b) the present result
appear to be consistent with $\lambda \sim 0.8$ and 
$\bar{\lambda} \sim 3.2$ (thus $\bar{\lambda}/\lambda \sim 4$)
being consistent with the scaling relation \eq{eq-lambda-relation} 
and the inequality $\bar{\lambda}/\lambda \geq  1$ \eq{eq-ratio-lambda}.
Indeed it can be seen that the recovery time at which the data 
merges with the reference data of $C_{0}(\tauh+\tw,\tw)$
(thus $\rho^{A} \to 1$)
is already as large as $O(10^{4})$ (MCS) even with very short
perturbation $\tau_{\rm p}=10$ (MCS).

The factorization in \eq{eq-c-memory} strongly suggests
independence of the evolution of the amplitude 
of the bias or the order parameter and size
of the domain. Consequently somewhat surprisingly the above result 
suggest the memory peaks can be {\it always} 
identified no matter how long the perturbation is kept on.  
Note that nothing special happens when $L(\taup)$ exceeds $L(\tw)$.
Only the amplitude of the signal will be smaller for 
longer perturbation so that higher resolution is required. 

Rather amusingly the factorization \eq{eq-c-memory} allows 
one to extract the growth of  the amplitude of the bias $\rho$ 
even with no knowledge of the underlying equilibrium state 
$\Gamma/\bar{\Gamma}$ thanks to the gauge (ghost) invariance
of the auto-correlation functions.
Probably it is interesting to apply the same trick to other 
spin-glass models. In  $d=3$ Ising EA model the data reported 
in \cite{kissanschrie96} on the relaxation of the auto-correlation 
function in isothermal aging suggest roughly $\bar{\lambda} \sim 2$
and hence $\bar{\lambda}/\lambda \sim 2$ assuming the scaling
relation \eq{eq-lambda-relation}.

{\bf Weakly perturbed regime-}
Naturally the factorization of the time evolution of the
bias and the size of the domains \eq{eq-c-memory} 
should also hold  in the weakly perturbed regime.
In the 4 dim EA model we also performed bond cycling simulations
operated in the weakly perturbed regime with very small $p$ such as
$p=0.02$. There we found the recovery of the bias
is {\it additive} as suggested by \eq{eq-rho-rec-weak} 
and the recovery time was found to be the trivial one
$\tau_{\rm rec} \sim \taup$ being consistent with \eq{eq-tau-rec-weak}.

\subsection{Magnetic susceptibilities}

In experiments AC/DC magnetic susceptibilities are often used to study
dynamics of spin-glass materials. As noted 
in section \ref{subsec-physobs} these are also
essentially gauge invariant quantities. Here we discuss possible scaling
properties of those in the healing stage. 

The relaxation of the AC susceptibility of frequency $\omega$ 
can be considered as a probe of the increase of the effective stiffness of 
a droplet excitation of size $L_{T}(1/\omega)$ 
due to decrease of domain wall density.\cite{fishus88noneq}
The scaling ansatz which follows
this picture has been supported by recent numerical and experimental
studies \cite{schetal93prb,komyostak2000,jonetal2002PRL,yoshuktak2002}.
In the healing stage, there must be excess contributions from the
domain walls around the islands of the minority phase. 
A given spin can be surrounded by such
a wall with probability 
$p_{\rm minor}(\tauh,\taup)=(1-\rho_{\rm rec}(\tauh,\rho_{\rm rem}(\taup)))$ 
(See \eq{eq-p-minority}) which increases with $\taup$ and 
decreases with $\tauh$.
 Then a natural scaling is
\be
\chi''(\omega, \tauh+\taup+\tw)= 
p_{\rm minor}(\tauh;\taup) \chi''_{0}(\omega,\tauh)+\chi''_{0}(\omega,\tauh+\tw)
\qquad \mbox{for} \qquad  p_{\rm minor} \ll 1 \qquad 
\label{eq-chi-rej-mem}
\ee
Here  $\chi''_{0}(\omega,t)$ is the AC susceptibility of isothermal 
aging starting from random initial condition at $t=0$.
The 1st term is the excess response due to the minority phase.
Since $p_{\rm minor}(\tauh,\taup)$ decreases with time $\tauh$,
the excess part slowly fades away. The 2nd term in the r.h.s.is due to the
outer-coarsening which is just the AC susceptibility without 
perturbation. 
If the cycling is operated in the weakly perturbed regime, the excess
part will fade away at the time scale $\tau_{\rm rec}^{\rm weak}$ given 
in \eq{eq-tau-rec-weak} while it will take an extremely 
long time $\tau_{\rm rec}^{\rm strong}$
given in \eq{eq-tau-rec} in the strongly perturbed regime.
It is interesting to note that anomalously large 
recovery time which apparently exceeds the simple 
estimate \eq{eq-tau-rec-weak},
which is valid only in the perturbative regime, has been found in recent 
measurements of the AC susceptibility \cite{sasetal2002,jonetal2003}.

In isothermal aging which start from a random initial condition at $t=0$
the ZFC susceptibility $M_{\rm ZFC}(\tau=t-\tw)$ measured under
a probing field switched on after waiting time $\tw$ 
exhibits a  rapid increase 
at around $\tau \sim \tw$. The latter is reflected 
as a peak of the relaxation rate $S(\tau)=d S(\tau)/d\log(\tau)$ 
at around $\tau \sim \tw$. \cite{lunetal83} 
In the cycling operated in the strongly perturbed regime, it is very
likely to happen that the population of the minority phase 
within the ghost domains $p_{\rm minor}(\tauh)$ 
remains non-zero at the time scale 
$\tauh \sim \tw$. This may explain the substantial 
reduction of the amplitude of the memory peak of $S(\tauh)$ at around 
$\tauh \sim \tw$ in one-step temperature-cycling experiments 
operated in the strongly perturbed regime \cite{grnetal90,jonetal2003}.

\section{Renormzalization of slow switching effects}
\label{sec-ren-heat-cool}

So far we considered idealized situations  that perturbations
are switched on/off instantaneously which is not
possible in reality. For example typically heating/cooling 
rates are $v_{T}=1$ Kelvin/second in ``quench''experiments 
\cite{jonetal98} which is equivalent to $v_{T}=10^{-15} J/{\rm MCS}$ 
in simulations (assuming $T_{\rm g}=10$ K and the microscopic 
time scale $\tau_{0}=10^{-13}$ (sec)). The surprising weakness 
of heating/cooling rate effect in spin-glasses \cite{jonetal98}
already suggests relevance of the chaos effects.

Let us illustrate here some important consequences of
such a slow switching by considering a continuous bond change protocol
as an example. Suppose that the signs of a fraction $p$ of 
$\pm J$ bonds in a temporally set ${\cal J}(t)$ are changed randomly
in a unit time $\tau_{0}$.
After some transients the system should become stationary by the chaos
effect such that the size of the ghost domains of $\Gamma^{{\cal J}(t)}$ 
becomes constant in time $L_{v_{J}}$ which decreases by increasing 
the bond change rate $v_{J}=pJ/\tau_{0}$.
Then we can consider for example a one-step bond cycling
$J_{A} (\tw) \to J_{B} (\taup) \to J_{A} (\tauh)$
with such a gradual bond changes. The point is that at length scales smaller than 
$L_{v_{J}}$ the whole cycling process just amounts to 
successive operations in the weakly perturbed regime.
There the multiplicative effects \eq{eq-tau-rec-multi} 
are avoided as explained in section \ref{subsec-weak}.
Then the cycling can be coarse-grained by taking $L_{v_{J}}$ 
as the new microscopic length scale instead of the overlap length
$\Lovlp$ between $A$ and $B$ which yields a coarse-grained 
cycling $J_{A} \to J_{B} \to J_{A}$
operated in a strongly perturbed regime with instantaneous 
bond changes. The scaling properties 
of the strongly perturbed regime will hold  for the latter but the 
original overlap length  $\Lovlp$ should be replaced by the 
{\it renormalized overlap length} $L_{v_{J}}$, for example in \eq{eq-tau-rec},
which leads to  a certain ``rounding'' of the strong chaos effects in realistic
circumstances. 

Although the temperature dependence of 
the growth law \eq{eq-growth} induce some obvious complications,
essentially the same argument for the case of continuous 
temperature changes leads to a corresponding renormalized overlap length 
$L_{v_T}$ which decreases with increasing heating/cooling rate $v_{T}$.

\section{Conclusion}

To summarize we studied how spin-glasses heal after being 
exposed to strong perturbations which induce the chaos effects
in simple perturbation-healing protocols (e.g. one-step temperature-cycling).
The bias or the order parameter within the 
ghost domains decays as $L^{-\bar\lambda}$ in the perturbation stage 
and increases as $L^{\lambda}$  in the healing stage
with increasing dynamical length scales $L$. The inequality of the exponents 
$\bar{\lambda} \geq  \lambda$  immediately suggests anomalously large 
recovery times of the order parameter. The memory auto-correlation function is 
suited for direct examination of the time evolution of the order parameter.
It should exhibit the memory peak in the healing stage 
at the time scale of the ``age'' imprinted in the ghost domains
due to the parallel evolution of the order parameter (inner-coarsening)
and the size of the ghost domains (outer-coarsening).
The predictions were checked quantitatively by numerical simulations
in  the 4 dim EA model.
It should be very interesting to measure experimentally 
the memory auto-correlation function by the 
noise-measurement technique \cite{HO02} for example in the standard one-step 
temperature-cycling protocol. 
Extensive experimental and numerical investigations 
which examine the ghost domain scenario
in other observables such as the AC/DC magnetic susceptibilities 
will be reported elsewhere \cite{jonetal2003}. 
Important features of the weakly perturbed
regime of the chaos effect were also discussed which leads to 
a proposal to take into account the effect of finiteness of 
switching on/off the perturbations in experimental circumstances
by the renormalized overlap length.

\ack The author thanks Jean Philippe Bouchaud, Koji Hukushima,
Petra E. J{\"o}nsson, Ana\"el Lema\^{\i}tre,  Philipp Maass, 
Roland Mathieu, Per Nordblad, Marta Sales, Falk Scheffler and Hajime
Takayama for
collaborations related to this work and useful discussions.

\vspace*{1cm}
\bibliographystyle{aip}
\bibliography{refs}

\begin{thebibliography}{10}
\bibitem{foot1}
Unfortunately the spherical model studied in \cite{yoslembou2001} is special in
  the sense that $\bar{\lambda}=\lambda=d/2$ so that $\bar{\lambda}/\lambda=1$.
  In the latter study, the possibility of the enormously large recovery time
  was not noticed.
\end{thebibliography}

\end{document}